\documentclass[aps,prl,twocolumn,showpacs]{revtex4}

\usepackage{epsfig}
\usepackage{graphicx}

\begin{document}

\DeclareGraphicsExtensions{.eps,.EPS}

\title{Accumulation and thermalization of cold atoms in a finite-depth magnetic trap}
\author{R. Chicireanu, Q. Beaufils, A. Pouderous, B. Laburthe-Tolra, E. Mar\'echal, J. V. Porto$^{\dag }$, L. Vernac, J. C. Keller, and O. Gorceix}
\affiliation{Laboratoire de Physique des Lasers, CNRS UMR 7538, Universit\'e Paris 13,
99 Avenue J.-B. Cl\'ement, 93430 Villetaneuse, France}
\affiliation{$^{\dag }$ National Institute of Standards and Technology, Gaithersburg, Maryland 20899, USA }

\begin{abstract}

We experimentally and theoretically study the continuous accumulation of cold atoms from a magneto-optical trap (MOT) into a finite depth trap, consisting in a magnetic quadrupole trap dressed by a radiofrequency (RF) field. Chromium atoms ($^{52}$Cr) in a MOT are continuously optically pumped by the MOT lasers to metastable dark states. In presence of a RF field, the temperature of the metastable atoms that remain magnetically trapped can be as low as 25 $\mu K$, with a density of $10^{17}$ atoms.m$^{-3}$, resulting in an increase of the phase-space density, still limited to 7.10$^{-6}$ by inelastic collisions. To investigate the thermalization issues in the truncated trap, we measure the free evaporation rate in the RF-truncated magnetic trap, and deduce the average elastic cross section for atoms in the $^5D_4$ metastable states, $\sigma_{el} = 7.0 \times 10^{-16} $m$^2$. We finally discuss the possibilities for using this scheme of continuous accumulation and RF evaporation to rapidly reach high phase-space densities from a MOT.

\end{abstract}


\date{\today}

\maketitle

Magneto-optical trapping is one of the greatest recent advances in atomic and molecular physics, opening many new areas in physics, including the study of Bose-Einstein condensation in dilute systems. One of the advantages of a magneto-optical trap (MOT), is that cooling and trapping are simultaneous. However, inherent limitations of the cooling mechanism, for example light-assisted collisions or multiple scattering of light, limit typical phase-space densities to $10^{-7}$ when strong fluorescence lines are involved in the cooling transition \cite{Townsend}. To reach higher phase-space densities, it is usually necessary to use sequential operations, such as cooling in molasses, loading in conservative traps, and forced evaporation, which greatly limits the rate at which a Bose-Einstein condensates (BEC) can be produced.

New possibilities arise for atoms whose electronic level structure includes metastable dark states (such as Sr, Er, Yb, or Cr). Spontaneous emission from the excited state of the cooling transition depumps atoms into a dark metastable state, which can be trapped, either optically, or magnetically. This provides an interesting means to continuously accumulate atoms in a magnetic trap \cite{pfauinelastique}, or for a continuous loading of an optical or magnetic waveguide. Cold metastable atoms are directly produced inside the waveguide or the trap, which reduces heating associated with the sequential loading from a MOT. This may have promising applications in the prospect of continuously producing a coherent beam of atoms, by performing forced evaporation as the atoms propagate along a waveguide \cite{Lahaye}.

\begin{figure}
\centering
\includegraphics[width= 2.8 in]{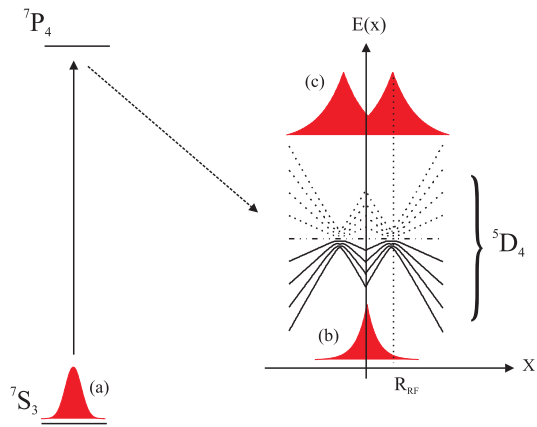}
\caption{\setlength{\baselineskip}{6pt} {\protect\scriptsize
Relevant level structure for chromium. The cooling transition is $^7S_3 \rightarrow ^7P_4$. Atoms are depumped to the $^5D_4$ metastable state. For this state, we also present a sketch of the RF dressed magnetic Zeeman eigen energies. Along the x axis, the avoided crossing occurs at $R_{RF}$. The density profiles are also qualitatively represented (shaded area) for atoms in the MOT (a), and for metastable states depending on the sign of their magnetic quantum number (b) and (c).}} \label{principe}
\end{figure}

In this paper, we study the accumulation of metastable chromium atoms in a magnetic trap (MT), dressed by a radiofrequency (RF) field (see Fig \ref{principe}) \cite{zobay} \cite{Colombe04}. As in \cite{pfauinelastique} and \cite{chicireanu06}, chromium atoms in the $^7S_3$ ground state are captured in a MOT, and depumped to dark metastable states $^5D_{4,3}$. We observe different regimes as a function of the RF frequency. When the RF frequency is small, atoms in internal states adiabatically connected to high field seeking states at small magnetic fields are accumulated in a (3D) W-shaped potential. When the RF frequency $\nu$ is larger than the MOT temperature $T_{MOT}$, i.e. $h \nu > k_B T_{MOT}$ (where $h$ is the Planck constant, and $k_B$ the Boltzmann constant), atoms in low-field seeking states are accumulated in a finite-depth potential. In this latter case, we achieve the continuous loading of a trap where evaporation is also continuous.

The purpose of this paper is to carefully study the accumulation of atoms in such a finite-depth trap, including the analysis of the different loss and heating mechanisms, the role of elastic collisions and the issues of evaporation and thermalization. We will show that fairly large phase-space densities can be reached. In the case of chromium, the gain in phase-space density is nevertheless limited by large inelastic loss rates associated with collisions with atoms from the MOT, and by large inelastic collision rates between metastable atoms.

In the first part of this paper, we describe our experimental observations, which show the fast (300 ms) production of a cloud whose phase-space density can reach 7.$10^{-6}$. This is about a fifty fold increase compared to what is typically obtained in standard MOTs, and a ten fold increase in the phase-space density compared to the Cr MOT we operate.

 In the second part, we present a theoretical model based on rate equations for the atom number and the total energy of the sample, to interpret our experimental results and determine which processes limit the phase-space density. This model includes loading and loss processes, and assumes that thermal equilibrium is reached on a timescale which is short compared to the timescales associated with loss processes. We also assume that the trap depth is large enough for the thermal distribution in the truncated trap to be identical to the thermal distribution in the non-truncated trap (\cite{evaporation}, \cite{Doyle}). The comparison between our experimental data and our simulation points towards a strong heating process due to collisions between the magnetically trapped atoms and the MOT atoms, which strongly limits the achieved phase-space density.

In the third part, we describe additional experiments, which allowed us to measure the evaporation rate, by comparing the loss rates and the heating rates with and without RF. From these measurements, we infer the average cross section for elastic collisions between metastable chromium atoms, which was never measured before. This measurement indicates that thermalization is not achieved when the atoms are accumulated in the truncated magnetic trap at low RF frequencies. We finally discuss the consequences of our observations in the prospect of reaching high phase-space densities by continuously accumulating atoms in a truncated trap.

\section{Continuous accumulation in a RF-dressed magnetic trap}

Our experimental setup was described in a previous paper \cite{chicireanu06}. Chromium atoms are heated to 1500 C, decelerated in a Zeeman slower, and captured in a MOT whose magnetic field gradient is 18 G/cm in the vertical direction, and 9 G/cm in the other two directions. Compared to reference \cite{chicireanu06}, the MOT laser beam $1/e^2$ radius is reduced to 3 mm. Here, we specifically study the accumulation of $^{52}$Cr atoms in the $^5D_4$ state, in presence of a RF field. We generate the RF field $B_z$ using a commercial synthesizer, a 11 dB pre-amplifier, and a 10 W amplifier, sending the RF power to a copper coil (2.5 turns, 10 cm diameter). After a given time $\tau$ ($=3s$ for most experiments reported in this paper) during which we run the MOT without any repumper, we shine the red repumping light (663 nm for $^5D_4 \rightarrow ^7P_3$) for 4 ms, and capture an image of the cloud, using dark ground absorption imaging along the horizontal $y$ axis (1:1 imaging, using a telescope with two 2 inch diameter achromatic doublet lenses, and a CCD camera), with a 100 $\mu$s resonant pulse well below optical saturation. Our imaging resolution is limited by the size of the camera pixels, 6.5 $\mu$m. The repumping time, 4 ms, is long enough for all atoms in the $^5$D$_4$ state to be repumped to the $^7$S$_3$ electronic ground state, but still short compared to all other timescales in the experiment. In addition, the recoil energy associated with repumping is negligible compared to the kinetic energy of the atoms. We therefore consider that the temperature and densities measured for atoms in the $^7$S$_3$ after repumping are equal to those of atoms in the $^5$D$_4$ state before repumping.

For in situ density measurements, the absorption pictures are taken in presence of the magnetic field gradient. We analyze the images by a fitting procedure which assumes thermal equilibrium in a trap potential $V(x,y,z)=m_S g_J \mu_B B' \sqrt{x^2+y^2+4z^2}$ ($m_S$ is the magnetic sublevel, $g_J = 1.5$ the Land\'e factor, $\mu_B$ the Bohr magneton, and $B'=9$ G/cm the magnetic field gradient). From this analysis, we first deduce the total atom number. The cloud size, typically 500 $\mu$m, is small enough that we can neglect the Zeeman effect on the cycling transition (the Zeeman shift is less than one tenth of the linewidth). Since the atomic sample is unpolarized and the magnetic field is inhomogeneous, we assume an average Clebsch Gordan coefficient of 3/7 for the resonant photon absorption cross section $\sigma_{abs} = \frac{3 \lambda^2}{2 \pi}$ (where $\lambda = 425.5$ nm is the photon wavelength), corresponding to an uniform magnetic sub-level distribution. From optical pumping calculations simulating an unpolarized sample, we estimate that the corresponding error bar on the number of atoms is $\pm 15$ percent: this is the main contribution to the error bar in the density and in the determination of chromium inelastic loss parameters described later in this paper.

\begin{figure}[h]
\centering
\includegraphics[width=2.8 in]{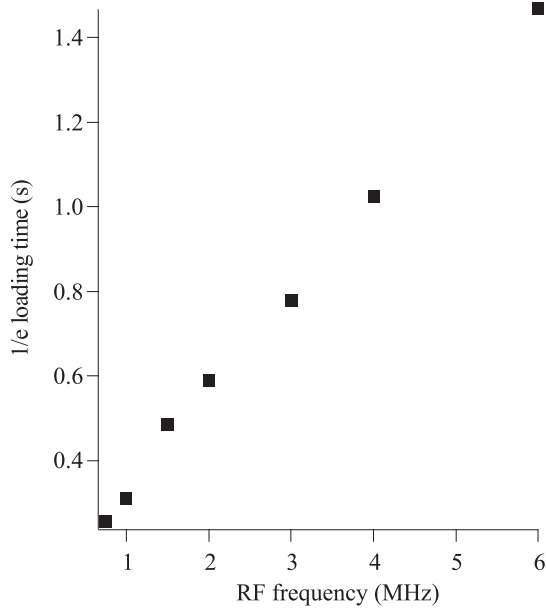}
\caption{\setlength{\baselineskip}{6pt} {\protect\scriptsize
Accumulation time in the truncated MT $T_{load}$, as a function of the RF frequency.}} \label{tempsaccumulation}
\end{figure}

\begin{figure}[h]
\centering
\includegraphics[width=2.8 in]{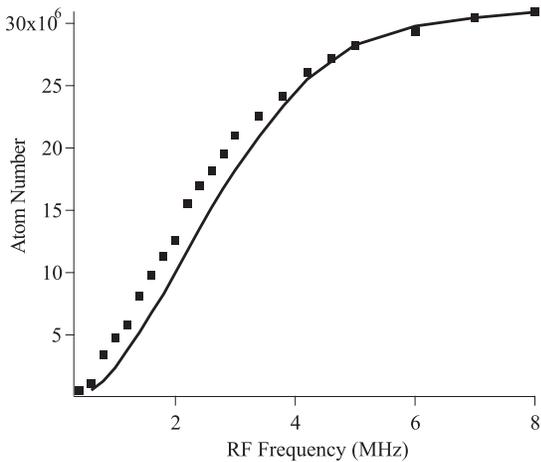}
\caption{\setlength{\baselineskip}{6pt} {\protect\scriptsize
Atom number after 3 s of accumulation, as a function of the RF frequency (squares). Solid line : theory (see text).}} \label{nombreatomes}
\end{figure}

We measure the number of atoms accumulated in the truncated MT, as a function of time $\tau$, for different RF frequencies $\nu$. At any frequency, we observe an exponential loading of the trap, and we plot in Fig \ref{tempsaccumulation} the $1/e$ accumulation time $T_{load}$ as a function of the RF frequency. The accumulation time increases approximately linearly from 0.3 s for $\nu = 1$ MHz to 1.4 s for $\nu > 6 MHz$. Without RF, the accumulation time is 1.4 s. In Fig \ref{nombreatomes}, we show the steady state atom number after 3 s of accumulation in the truncated MT as a function of the RF frequency.

\begin{figure}[h]
\centering
\includegraphics[width=2.8 in]{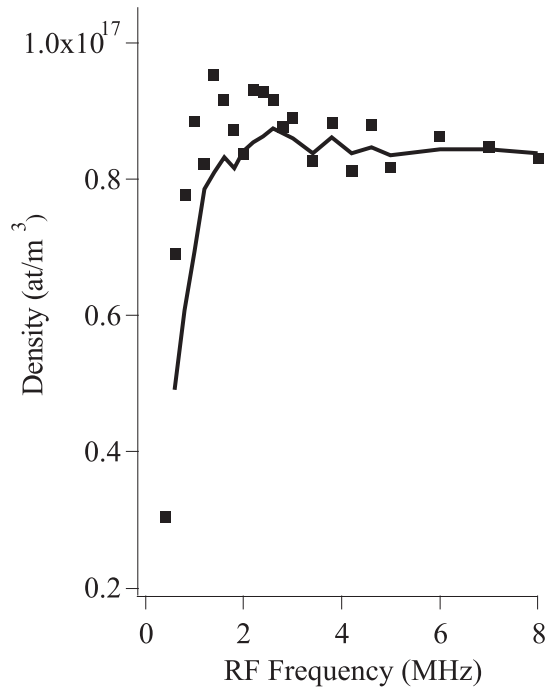}
\caption{\setlength{\baselineskip}{6pt} {\protect\scriptsize
Peak atom density after 3 s of accumulation (squares). Solid line : theory (see text).}} \label{densite}
\end{figure}

Since the atom cloud is not polarized, the exact analysis of the absorption images is complicated. For sake of simplicity, we will assume in this paper that all atoms experience an average potential $\overline{m_S} g_J \mu_B B' \sqrt{x^2+y^2+4z^2}$, where $\overline{m_S}$ is an average magnetic quantum number, which we deduce from direct temperature measurements. The density profile at thermal equilibrium is $n_{MT}=n_0 exp(-V_0 /k_BT \sqrt{x^2+y^2+4z^2})$, where $T$ is the temperature, and $V_0= \overline{m_S} g_J \mu_B B'$. From the analysis of the absorption images, we can deduce the $1/e$ radius of the cloud $d$, as well as the peak atom density $n_0 = \frac{7}{6} \frac{O.D.}{d \sigma_{abs}}$, where $O.D.$ is the experimental central optical depth. After 3 s of accumulation, the observed peak atom density is almost independent of the RF frequency, except when this frequency is set below approximately 1 MHz (see Fig \ref{densite}).

In addition to density measurements, we perform direct temperature measurement, by analyzing the free-fall expansion of the atoms after release from the magnetic trap. The current in the coils producing the magnetic field gradient is switched off in ~500 $\mu s$. Eddy currents generate a non-negligible bias magnetic field at the atom location for more than 5 ms, but this does not modify the ballistic expansion of the atoms. To deduce the temperature, we therefore analyze times of flight larger than 600 $\mu$s, by fitting the integral of the absorption images by gaussian functions. Although the initial shape of the cloud is not gaussian, we numerically checked that the systematic error bar associated to this procedure is less than 10 percent, when fitting the vertical direction (along which the extension of the cloud is the smallest), for temperatures on the order of 60 $\mu$K.

\begin{figure}[h]
\centering
\includegraphics[width=2.8 in]{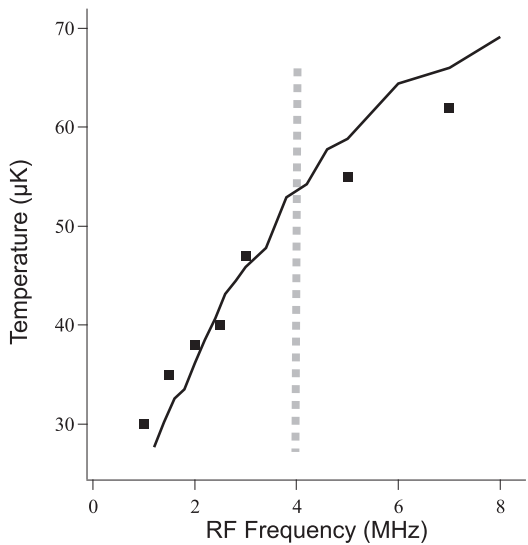}
\caption{\setlength{\baselineskip}{6pt} {\protect\scriptsize
Temperature of the atoms after 3s of accumulation, as a function of the RF frequency. The experimental error bar on the temperature is about 10 $\mu$K. Solid line : theory (see text). Experimental points at the left of the vertical dotted line correspond to a situation where thermal equilibrium is not reached (see text).}} \label{temperature}
\end{figure}

Using this temperature and the measured $1/e$ size of the cloud in situ $d$, we deduce the mean magnetic quantum number $\overline{m_S} = k_B T / d g_J \mu_B B' $ of the atoms in the MT. Interestingly, $\overline{m_S}$ does not change much with the RF frequency, staying close to 2.

\begin{figure}[h]
\centering
\includegraphics[width=2.8 in]{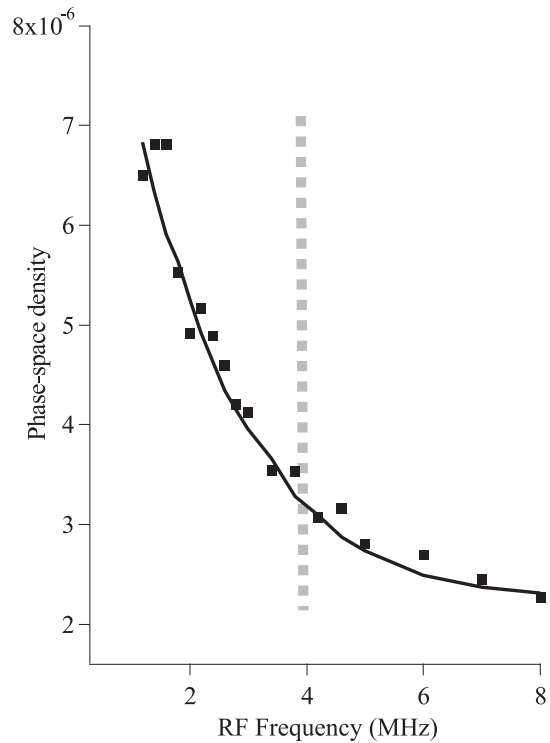}
\caption{\setlength{\baselineskip}{6pt} {\protect\scriptsize
Phase-space densities after 3 s of accumulation in the RF trap, as a function of the RF frequency (Squares). Solid line: theory. Experimental points at the left of the vertical dotted line correspond to a situation where thermal equilibrium is not reached (see text).}} \label{figurephasespace}
\end{figure}

In figure \ref{figurephasespace}, we show the steady state phase-space density after $\tau = $3 s of accumulation, as a function of the RF frequency. Because the measured temperature of the atoms decreases with decreasing RF frequency (Fig \ref{temperature}), and because the measured peak atom density remains almost constant (Fig \ref{densite}), the phase-space density increases at low RF frequencies (Fig \ref{figurephasespace}), to reach a maximum value of $(7.0 \pm 1.5)$  $10^{-6}$ at $\nu = 1.5$ MHz. The error bar is due to the systematic errors on the density and on the temperature measurements (see above). The maximum phase-space density value is significantly higher than typical phase-space densities achieved in a MOT involving a strong resonance line.

For frequencies lower than 1.5 MHz, this trend inverts: when lowering the RF frequency, the density of the atoms decreases (see Fig \ref{densite}), and we infer a decrease of the phase-space density, instead of an increase. For such low RF frequencies, the atoms in the low-field seeking states see a trap whose depth is lower than the MOT temperature, and most of them are therefore not trapped. However, for frequencies on the order of 800 kHz or below, new interesting features appear in the absorption images. The $1/e$ size of the imaged cloud becomes much larger, and the atom lifetime is longer than the lifetime measured at 1 MHz. We interpret these pictures as resulting from atoms trapped in the W-shaped adiabatic potential corresponding to states, dressed by the RF field, adiabatically connected to high-field-seeking states at B=0 \cite{zobay},\cite{Colombe04}. Indeed, the cloud size increases linearly with RF frequency, and its vertical position also shifts linearly with RF frequency, as is expected in the RF-dressed potential, whose size is determined by $R_{RF} = h \nu_{RF} / g_J \mu_B B'$ (see Fig. (\ref{principe})). For low RF frequencies, we found that the number of atoms increases before decreasing for large RF power. We attribute this effect to higher harmonic frequencies appearing at high RF power : a substantial power of the second harmonics of the RF frequency is then exciting the atom sample, so that the atoms reaching $2 R_{RF}$ have a large probability of being extracted from the dressed trap.

In the remaining of this paper, we will focus on RF frequencies larger than 1 MHz, for which we only trap low-field seeking states. One of the main goals of this paper is to quantitatively understand the achieved phase-space densities in this regime. In the next part, we describe a theoretical model, which assumes that the elastic scattering rate is high enough to insure thermal equilibrium. Then, in Part 3, we will experimentally determine the elastic scattering rate, in order to determine in which RF regime this theoretical model is expected to be valid.

\section{Theoretical model and interpretation}

Our theoretical model for the accumulation of the atoms in the RF-truncated magnetic trap is based on rate equations for the number of atoms and the total energy of the system. It is elaborated from \cite{evaporation} and \cite{comparat06}. We include in our model the density dependent loss processes, assuming that the density profile of the atoms is, at all time, well described by the Boltzmann distribution at thermal equilibrium in the non-truncated trap. This relies on two main assumptions: first, the elastic collision rate is large enough to reach thermal equilibrium on the timescale of the experiment (defined by the dominant loss process); second, the trap depth is large enough compared to the magnetic trap temperature, so that one can neglect modifications in the thermal distribution linked to the finite depth of the potential (\cite{evaporation}, \cite{Doyle}). In our experiment, thermal equilibrium is most likely not reached at the beginning of the loading of the MT, when the number of atoms is still small. For this reason, we do not expect this model to describe the dynamics of the loading. However, the steady state reached at the end of the accumulation is the one predicted by our model, provided the rate of elastic collisions in the $^5$D$_4$ state is high enough (or if the steady state density profile is close enough to the Boltzmann density distribution).

To be able to capture the physics of the accumulation in the truncated MT, our model takes into account the loading of the atoms and various loss mechanisms. The key point for the loading parameters is that the loading rate and the average energy per atom loaded in the MT both depend on the RF frequency. We experimentally checked that the MOT atom number is not modified by the presence of the RF field. We therefore assume that the production rate of atoms in the metastable states does not depend on the RF frequency $\nu$. However, once the atoms decay into the $^5$D$_4$ state, we assume that they are only trapped if their total energy is smaller than the trap depth. The loading rate $\Gamma(\nu)$ is therefore obtained by multiplying $\Gamma$, the loading rate without RF, by the probability $P(\nu )$ that an atom transferred from the MOT - at a position $(x,y,z)$, and with a velocity $v$ - into the metastable state has a total energy $E_{tot}=1/2mv^{2}+V_{0}\sqrt{x^{2}+y^{2}+4z^{2}}$ smaller than the depth of the truncated MT, $\overline{m_{S}}h\upsilon $:

\begin{equation}
\Gamma (\nu )=P(\nu )\Gamma
\end{equation}%

$P(\nu )$ can be evaluated numerically by integrating the phase-space density over the domain $D$ defined by the condition $E_{tot}<\overline{m_{S}}h\upsilon :$

\begin{equation}
P(\nu )=\frac{\int\limits_{D}du d^{3}\overrightarrow{r}\sqrt{u%
}e^{-u}e^{\frac{-(x^{2}+y^{2}+4z^{2})}{2w_{_{MOT}}^{2}}}}{\int\limits_{R^{4}} du d^{3}%
\overrightarrow{r}\sqrt{u}e^{-u}e^{\frac{-(x^{2}+y^{2}+4z^{2})}{2w_{MOT}^{2}}}}
\end{equation}

In this formula, we have used the fact that the phase-space density is proportional to the spatial density in the MOT, $n_{MOT}(\overrightarrow{r})=n_{0,MOT}\exp (-x^{2}+y^{2}+4z^{2})/2w_{MOT}^{2})$, and to the $\sqrt{u}\exp(-u)$ factor coming from the free space kinetic energy distribution. This MOT density mostly reproduces the experimental one, and allows great mathematical simplifications. We found in our simulations that the exact form of the MOT density mainly affects the loading parameters and the results of our simulations at low RF frequencies, but doesn't modify much the inelastic loss parameters that we deduce from the simulations.

We also calculate the average total energy per atom trapped in the truncated MT, $E(\nu )$:

\begin{equation}
\frac{E(\nu )}{k_{B}T_{MOT}}=\frac{\int\limits_{D}du d^{3}\overrightarrow{r}(u+ \frac{V_{0}\sqrt{%
x^{2}+y^{2}+4z^{2}}}{k_{B}T_{MOT}})%
\sqrt{u}e^{-u}e^{\frac{-(x^{2}+y^{2}+4z^{2})}{2w_{_{MOT}}^{2}}}}{\int\limits_{R^{4}} du d^{3}%
\overrightarrow{r}\sqrt{u}e^{-u}e^{\frac{-(x^{2}+y^{2}+4z^{2})}{2w_{MOT}^{2}}}}
\end{equation}

Our model includes different loss mechanisms:

(i) collisions with the background gas, mostly hot atoms coming from the oven (rate $\Gamma_0$);

(ii) losses corresponding to inelastic collisions with atoms from the MOT, with a rate

\begin{eqnarray}
\Gamma_1=1/N \int d^3 r \beta_{PD} n_{MOT}(\vec{r}) n_{MT}(\vec{r})
\end{eqnarray}

where $N$ is the total number of atoms in the magnetic trap, and $\beta_{PD}$ is the inelastic loss parameter for collisions between $D$ atoms and $P$ atoms;

(iii) two-body losses (rate $\Gamma_2 \equiv L_2 n_0 / 2^3$, where $n_0$ is the MT peak atom density);

(iv) three-body losses (rate $\Gamma_3 \equiv L_3 n_0^2/3^3$. In general $L_3$ has to be measured. However, in the limit of the existence of a weakly bound dimer, $L_3$ is related to the scattering length of the atoms to the fourth power \cite{Fedichev});

(v) Majorana losses (rate $\Gamma_{maj}=C \frac{\mu}{\Delta \mu} \frac{\hbar}{md^2}$, where $\mu$ is the atom magnetic moment, $\Delta \mu$ is the change in magnetic moment due to the Majorana spin flip, C is a numerical factor, on the order of 3, and $d$ is the 1/e radius of the cloud \cite{petrich95});

(vi) Evaporation related to elastic collisions, with rate $\Gamma_{ev}$:

\begin{eqnarray}
\Gamma_{ev} = \frac{\sqrt{2}}{8} \Gamma_{el} \times f(\eta)
\label{evapcoll}
\end{eqnarray}

where $\Gamma_{el} = n_0 \sigma_{el} \bar{v}$ is the elastic collision rate, with $\sigma_{el}$ the elastic cross section and $\bar{v} = \sqrt{\frac{8k_BT}{ \pi m}}$ the mean thermal velocity. $\eta$ is the ratio of the trap depth to the trapped atom temperature, and $f(\eta)$ is given in equation (11) of ref \cite{Doyle} (which is valid for $\eta \geq 4$).

We neglect any elastic collisions between atoms in the metastable states and atoms from the MOT \cite{notethermalisationMOT}. The atom number and total energy rate equations read:

\begin{eqnarray}
\frac{dN}{dt} = \Gamma(\nu)-(\Gamma_0+\Gamma_1+\Gamma_2+\Gamma_3+\Gamma_{maj}+\Gamma_{ev})N
\label{evol}
\end{eqnarray}

\begin{eqnarray}
\frac{d(9/2NT)}{dt} = E(\nu)\Gamma(\nu) -\sum_{i=0,1,2,3,maj,ev} f_i \Gamma_i NT
\label{eqevap}
\end{eqnarray}

where the $f_i k_B T$'s describe the average energy loss when one atom is lost by the corresponding mechanism. In a linear trap, one shows that $f_0=9/2$, $f_2=3$, $f_3=5/2$, $f_{maj}=5/2$. For instance, to evaluate $f_2$, one needs to integrate over the whole cloud the local inelastic rate $L_2 n^2_{MT}(r)$ times the local mean energy of the atoms $V(r)+3/2 k_B T$:

\begin{eqnarray}
f_2 \Gamma_2 NT = - L_2 \int d^3 r n_{MT}^2(r)\left(V(r)+3/2 k_B T\right)
\end{eqnarray}

If the trap is deep enough, $f_{ev}=\eta+1$ (see \cite{Doyle}). In addition,

\begin{eqnarray}
f_1 \Gamma_1 NT = \int d^3 r \beta_{PD} n_{MOT}(\vec{r}) n_{MT}(\vec{r})(V(r)+3/2 k_B T)
\end{eqnarray}.

We stress that there are few adjustable parameters, since $n_{0,MOT} = 7.5 \times 10^{16}$ m$^{-3}$,  $T_{MOT} =$ 120 $\mu$K and $w_{MOT}=$ 100 $\mu$m are independently measured. As shown below, we determine $L_2$ from an independent measurement. $\Gamma_3$ and $\Gamma_{ev}$ both depend on the scattering length $a$, which is used as a free parameter here. $\Gamma$ and $\beta_{PD}$ are the two other free parameters.

We numerically solve equations (\ref{evol}) and (\ref{eqevap}) using the experimentally measured $\overline{m_{S}}$ for each RF frequency. We first fit our numerical results to the experimental results for large RF frequencies ($\nu > 4$ MHz). In this regime, we expect that evaporation is negligible, because the trap depth is much larger than the temperature of the atoms: for instance, a RF frequency of 8 MHz corresponds to $\eta > 10$, so that the evaporation time is much longer than the accumulation time in the truncated trap. Therefore, we set $\Gamma_{ev}=0$ for the comparison to the experiment at large RF frequencies. The comparison to our experimental results for the temperature and number of atoms after 3 s of accumulation set the parameters $\Gamma = 3.3 \times 10^7$ $s^{-1}$ and $\beta_{PD} = (4.9 \pm 0.3 \pm 0.5) \times 10^{-10}$ $cm^3s^{-1}$. The error bars are respectively the statistic error bar of the fit, and the systematic error bar associated to the uncertainty on the total atom number. With these values, the agreement between theory and experiment is satisfactory, for RF frequencies exceeding 4 MHz (see Fig \ref{nombreatomes}, \ref{densite}, \ref{temperature} and \ref{figurephasespace}). We note that the value of $\beta_{PD}$ that we deduce from our analysis is close to what was measured in reference \cite{clip}.

For large RF frequencies, we checked that evaporation, Majorana losses, and three-body losses are negligible. The only other parameter which is not negligible is the inelastic two-body loss parameter, which is measured independently in our experiment (see below). The main conclusion of our analysis at high RF frequencies is that inelastic collisions with atoms from the MOT limit the total number of atoms that are accumulated in the MT, and produce a strong heating: the temperature of the atoms in the MT, which, in absence of collisions, could be as low at $T_{MOT}/3$ \cite{pfauinelastique}, raises to about $100$ $\mu$K for large RF frequencies. This drastically limits the achieved steady state phase-space density.

At RF frequencies lower than 4 MHz, the role of evaporation is not negligible. However, we find that varying the elastic cross section $\sigma_{el}$ does not lead to large modifications of the numerical results, so that it is in practice impossible to deduce from a fit a precise value of the elastic scattering rate. In addition, for the lowest frequencies, our model, which relies on the assumption that the gas is thermalized to the Boltzmann distribution, may not be fully applicable. The loading time $T_{load}$ of the atoms in the truncated MT gets smaller and smaller as the RF frequency decreases (see Fig. \ref{tempsaccumulation}): at low RF frequencies, the spatial overlap between the MT and the MOT gets better and better, which increases the loss rate related to inelastic collisions with atoms from the MOT (process (ii)), and reduces $T_{load}$. Therefore, Boltzmann thermal equilibrium is not necessarily reached before the steady state. This will be studied in detail in the following section.

\begin{figure}[h]
\centering
\includegraphics[width=2.8 in]{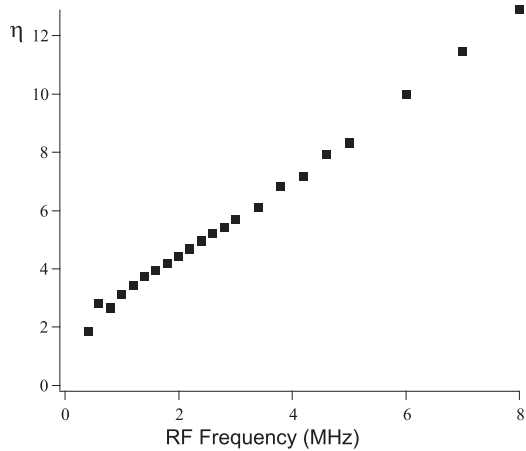}
\caption{\setlength{\baselineskip}{6pt} {\protect\scriptsize
$\eta$ as a function of the RF frequency. For $\eta \leq 4$, the density distribution at thermal equilibrium is modified by the trap truncation, which is not taken into account in the theoretical model.}} \label{eta}
\end{figure}

Effects linked to the small trap depth (and their consequences on the thermal distribution) may also have an important role at low RF frequencies. Our model, assuming that the thermal distribution in the truncated trap is identical to the one in a infinite depth potential, will only be valid provided $\eta \geq 4$ \cite{Doyle}. Although the sample is not polarized, the measurement of the $1/e$ radius of the cloud $d$ provides a good estimate for $\eta$, the ratio of the trap depth ($\overline{m_S} h \nu$) to the trap temperature $k_B T = V_0 d$, because this ratio does not depend on $\overline{m_S}$. We plot $\eta$ in Fig \ref{eta}, and find that our theoretical model may not be applicable when the RF frequency is smaller than about 2 MHz.

\section{Evaporative cooling and thermalization in a RF-truncated magnetic trap}

To investigate thermalization issues, we perform a new set of experiments to measure the lifetime and the heating rate of atoms in the $^5 D_4$ state with and without the RF field. For these experiments, the atoms are first accumulated without RF for 3 s in the magnetic trap. Then, the MOT is switched off, and the number of atoms and temperature of the cloud are measured after an adjustable delay $t$. This second step of the experiment is first realized without RF, then in presence of a 3 MHz RF field.

\begin{figure}[h]
\centering
\includegraphics[width=2.8 in]{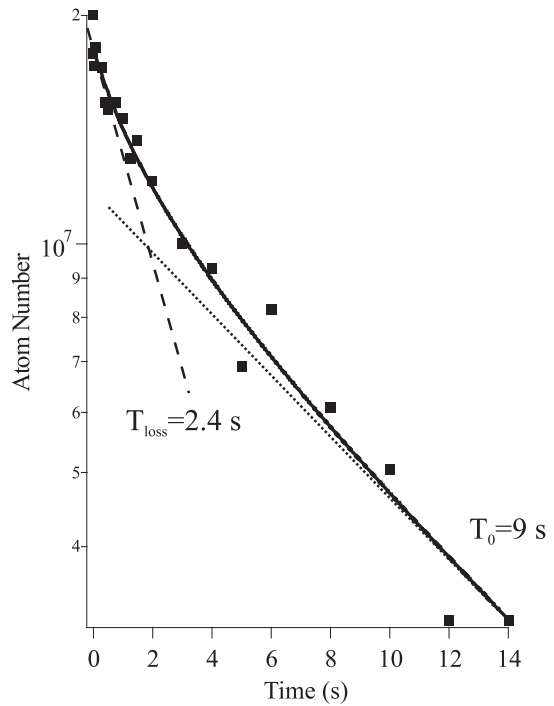}
\caption{\setlength{\baselineskip}{6pt} {\protect\scriptsize
Decay of atoms in the D state (squares). Solid line: result of the fit (see text). The dotted line is a result of a purely exponential fit of the experimental data for t$>5$s. The dashed line is a result of an exponential fit for t$<1s$.}} \label{inelastiquedecay}
\end{figure}

When no RF field is applied, we measure the decay of the metastable atoms in the MT (see Fig \ref{inelastiquedecay}), from which we deduce the inelastic loss parameter $L_2$ (see Fig \ref{inelastiquedecay}, which shows a clear non-exponential decay). We fitted the data of Fig \ref{inelastiquedecay} assuming, in addition to a one-body loss, a two-body loss parameter. Given the low atomic density, we assume that three-body processes are negligible. During the decay, the cloud volume increases approximately linearly. In the fit of Fig \ref{inelastiquedecay}, we assume a linear dependence to the volume on time, based on a fit to the measured time-dependent size. We found the inelastic loss parameter to be $L_2 = (3.3 \pm 0.5 \pm 0.5) \times 10^{-11} $ cm$^{3}$s$^{-1}$ (in agreement with \cite{clip}). The respective error bars are statistic error bar of the fit, and systematic error, mostly coming from the uncertainty on the total number of atoms. We stress that this inelastic loss parameter is much larger than the expected dipolar relaxation rate, which rules out dipolar relaxation as the main inelastic loss channel \cite{dipolarrelaxation}.

\begin{figure}[h]
\centering
\includegraphics[width=2.8 in]{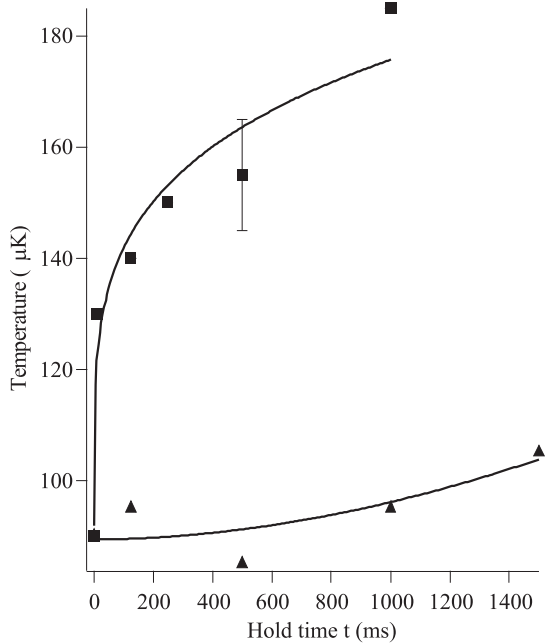}
\caption{\setlength{\baselineskip}{6pt} {\protect\scriptsize
Time evolution of the temperature without RF (squares) and with a $\nu =$ 3 MHz RF field (triangles). Lines are guides for the eye. The error bar illustrate the typical combined statistical and systematic errors on temperature measurements.}} \label{heatingornoheating}
\end{figure}

We repeat this lifetime experiment in presence of a 3 MHz RF field during the decay. Then, the decay of the cloud is modified by evaporation, which changes both the total loss rate, and the evolution of the temperature (see Fig \ref{heatingornoheating}). Our goal is to use these observations to estimate the evaporation rate and infer the elastic collision rate. However, one must first evaluate the modification of the lifetime of the cloud linked to spilling of atoms out of the trapping volume set by the RF frequency: the experimentally observed heating produces atoms whose energy is larger than the trap depth. They then leave the trap, without undergoing elastic collision, assuming that the mean free path of the atoms is larger than the size of the sample. Heating without RF translates into spilling when RF is applied, and the rate at which spilling occurs is therefore linked to the heating rate without RF. We measure the time evolution of the $1/e$ radius of the cloud $d$ without RF to estimate $\Gamma_{spill} \approx 0.1$ s$^{-1}$, the rate of atoms leaving the trapping volume set by the RF frequency, i.e. the rate of atoms spilled out of the trap volume.

From the atom number decay at short times, we deduce the experimental values of the atom loss rates without RF (see Fig \ref{inelastiquedecay}), $\Gamma_{loss} = 1/T_{loss} = 1/2.4$ s$^{-1}$, and with RF, $\Gamma_{loss, RF} = 1/1.35$ s$^{-1}$. We can therefore estimate the value of $\Gamma_{ev} + \Gamma_{spill} =\Gamma_{loss,RF}-\Gamma_{loss} = 0.32$ $s^{-1}$. From this equality, we obtain a first estimate for the evaporation rate from the modification in the lifetime of the cloud in presence of RF: $\Gamma_{ev} \approx 0.22 $ s$^{-1}$.

We deduce an additional independent estimate for the evaporation rate by comparing the heating rate in the RF truncated trap to the heating rate without RF. With the 3 MHz RF field on, we observe no substantial heating, as proven by temperature measurements as the hold time $t$ is varied (see Fig \ref{heatingornoheating}). This indicates that the cooling rate due to evaporation balances the heating rate related to two-body inelastic collisions. For RF frequencies higher than 3 MHz, the cloud does heat, whereas we observe cooling when the RF frequency is lower. This observation allows us to make a second determination of $\Gamma_{ev}$. Using (\ref{eqevap}), assuming only one-body, two-body losses and evaporation, and taking into account the experimental observation that $\frac{dT}{dt} \approx 0$, we have:

\begin{eqnarray}
\frac{9}{2}\frac{dN}{dt} T \approx -f_0NT \Gamma_0 -f_2NT \Gamma_2 -(\eta+1)\Gamma_{ev}NT
\end{eqnarray}

from which we deduce, using eq.(\ref{evol}):

\begin{eqnarray}
\Gamma_{ev} \approx \frac{(9/2-f_2)\Gamma_2}{\eta+1-9/2}
\label{noheating}
\end{eqnarray}

We estimate $\Gamma_{2,t=0} = \Gamma_{loss} - \Gamma_0$, where $\Gamma_0 = 1/T_0 = 1/9$ $s^{-1}$ is the loss rate at large time delays $t$, for which the atom number is small enough to neglect inelastic collisions (see Fig \ref{inelastiquedecay}). We attribute $\Gamma_0$ to the effect of collisions with the background gas (mostly hot atoms coming from the oven).  The $1/e$ size of the cloud and the experimentally measured temperature give $\overline{m_S} \approx 3.5$, and an average trap depth to trap temperature ratio of $\eta \approx 5.5 \pm 1$. We therefore deduce from eq. (\ref{noheating}) that $\Gamma_{ev}= 0,2 \pm 0.04 $ s$^{-1}$, in reasonable agreement with our first estimate.

From these measurements, we deduce the elastic collision rate, using eq. (\ref{evapcoll}) and eq. (11) of ref \cite{Doyle} to estimate the evaporation fraction $f(\eta)$: $\Gamma_{el} = $ 20 $\pm$ 4 $\pm$ 11 s$^{-1}$ (error bars respectively correspond to statistical error on $\Gamma_{ev}$ and to the estimated systematic error on $\eta$). The experimentally measured temperature is 100 $\mu K$, and the peak density $10^{11}$ cm$^{-3}$, at $t=0$. We therefore infer an average elastic cross section of $\sigma_{el}$ =(7.0 $\pm$ 1.4 $\pm$ 3.5 $)\times 10^{-16}$ m$^{2}$, close to the unitary limit at this temperature. The error bars respectively correspond to the statistical error, and the estimated systematic error (dominated by the estimate of $\eta$). This is, to our knowledge, the first measurement of an elastic cross section for a transition element in a metastable state. It should be mentioned that, in relation with the prospect for Bose-Einstein condensation with alkaline earth elements, the elastic cross section was also recently determined in metastable Ca \cite{Hansen}.

Using this average collision rate, we now check if our theoretical model can be used to analyze the evaporation during the decay, i.e. if thermal equilibrium is achieved throughout the decay, so that we can use equation (\ref{noheating}). To be able to describe the cloud by the Boltzmann distribution throughout the decay, it is necessary that a sufficient number of elastic collisions take place during the typical evolution time of the cloud. It is  difficult to define exactly a timescale to reach thermal equilibrium, as relaxation to the Boltzmann distribution is not exponential when this relaxation is driven by intraparticles collisions (as can be seen for example from eq. (14) in \cite{evaporation}). We perform numerical simulations of the Boltzmann equation, and find that, in practice, a ratio of about 4 elastic collisions per inelastic collision is enough to maintain the energy distribution of the most populated states close to a Boltzmann distribution. In our experiment, $\frac{\Gamma_{el}}{\Gamma_2} \approx \frac{20}{1/3.3} \approx 66$. The distribution should therefore be close to the Boltzmann distribution, and our determination of the elastic scattering rate in the $^5D_4$ state is therefore self-consistent.

\begin{figure}[h]
\centering
\includegraphics[width=2.8 in]{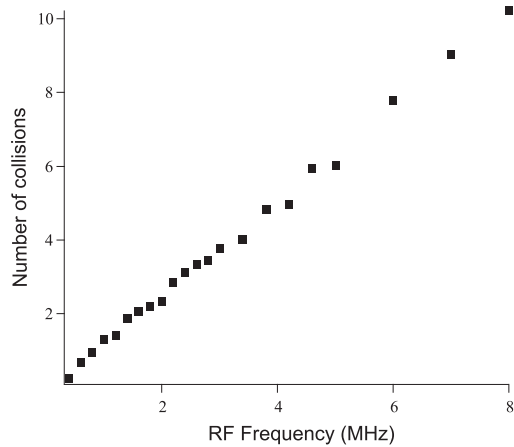}
\caption{\setlength{\baselineskip}{6pt} {\protect\scriptsize
Number of elastic collision before the steady state is reached when atoms are accumulated in the RF-truncated magnetic trap. When this number is smaller than approximately 4, thermal equilibrium is not reached, and our theoretical model does not apply.}} \label{numbercoll}
\end{figure}

We now can turn back to the analysis of the experiments described in the first part of this paper, and to the issue of thermal equilibrium in these experiments. For the continuous accumulation in the RF truncated trap, the loading time is mostly set by inelastic collisions with the MOT atoms. For low RF frequencies, for which the spatial overlap between the MT and the MOT is the best, the loading time $T_{load}$ is short, between 300 ms and 1s (see Fig \ref{tempsaccumulation}). Therefore, the elastic scattering rate is too small to provide thermal equilibrium at the end of the loading process. In Fig \ref{numbercoll}, we plot the approximate number of collisions before the steady state of accumulation is reached, qualitatively given by $n_{MT} \sigma_{el}\bar{v} T_{load} /2 $. From Fig. \ref{numbercoll}, we expect a lack of thermalization at RF frequencies below 4 MHz.

Preliminary numerical simulations based on the Boltzmann equation show that indeed thermalization is not achieved at low RF frequencies. For higher RF frequencies, these simulations also show that thermalization is not perfect. Since thermalization is not achieved in the experiment, especially at low RF frequency, the agreement between experimental results and theory may look surprisingly good. Such a good agreement relies on the fact that, at each RF frequency, the experimental density profile is always well fitted by a density profile assuming thermal equilibrium in a potential $V_0 \sqrt{x^2+y^2+4z^2}$. Therefore all loss terms in the system are well described by the terms $\Gamma_i$ and $f_i \Gamma_i$ in equations (\ref{evol}) and (\ref{eqevap}), which therefore still describe the experimental system, even if thermal equilibrium is not reached. For these reasons, the measurements of the inelastic loss parameters of chromium given in this paper should be correct. Furthermore, even if technically temperature and phase-space densities are not defined at the left of the vertical dotted line in Fig \ref{temperature} and \ref{figurephasespace}, the quantities we plotted still have a relevant physical meaning: we measured the spatial densities and the total energy of the system, which will define the thermodynamics quantities of the sample, once thermal equilibrium is reached.

In this paper, we have studied the accumulation of metastable atoms in a magnetic trap of finite depth. We demonstrated that phase-space densities 50 times larger than the typical ones in a MOT can be reached in less than 1s. However, we showed that inelastic collisions with atoms from the MOT reduce the loading time of the atoms in the trap so much that thermalization is not reached when the trap depth is too small. In addition, the heating rate associated with these inelastic collisions greatly reduce the achievable phase-space densities. This feature may be specific to chromium. With atoms for which such a loss process is reduced, we suggest that the continuous accumulation of atoms in a finite depth trap (either a magnetic trap or a dipole trap) could be  an efficient way to rapidly reach high phase-space densities. Finally, we also studied the inelastic collision properties of metastable chromium atoms, as well as their elastic properties, providing the first measurement of the elastic scattering rate for chromium atoms in the $^5D_4$ state. The ratio of the elastic rate to the inelastic rate is close to five, ruling out traditional forced evaporation techniques to reach BEC in these states.

Acknowledgements: LPL is Unit\'e Mixte (UMR 7538) of CNRS and of Universit\'e Paris Nord. We acknowledge financial support from Conseil R\'{e}gional d'Ile-de-France (Contrat Sesame), Minist\`{e}re de l'Education, de l'Enseignement Sup\'{e}rieur et de la Recherche, European Union (FEDER -Objectif 2), and IFRAF (Institut Francilen de Recherche sur les Atomes Froids).


\begin{thebibliography}{99}
\bibitem{notethermalisationMOT} The elastic cross section between atoms of the MOT and atoms of the MT is not known. However, even assuming that it reaches the unitary limit, the thermalization rate of the MT by the MOT atoms is always small compared to $\Gamma_1$
\end{thebibliography}
\end{document}